\documentstyle[12pt]{article}
\newcommand{\beq}{\begin{equation}}
\newcommand{\eeq}{\end{equation}}
\newcommand{\bea}{\begin{eqnarray}}
\newcommand{\eea}{\end{eqnarray}}
\newcommand{\pder}[2]{\frac{\partial #1}{\partial #2}}

\title{Interaction of surface waves with vorticity in shallow water}

\author{Enrique Cerda and
Fernando Lund \\Departamento de F\'\i sica \\ Facultad de Ciencias
 F\'\i sicas y Matem\'aticas \\ Universidad de Chile \\ Casilla 487-3,
 Santiago, Chile}

\date{}

\begin{document}

\maketitle

\paragraph{Abstract}
Vortical flows in shallow water interact with long surface waves by
virtue of the nonlinear terms of the fluid equations.
Analytical formulae are derived that quantify the spontaneous
generation of such
waves by unsteady vorticity as well as the scattering of surface waves
by vorticity. In a first Born approximation the radiated surface
elevation is linearly related to the Fourier transform of the vorticity.
The ``dislocated'' wavefronts that are analogous to the Aharonov-Bohm
effect are obtained as a special case.

\vspace{4em}
\begin{flushleft}
\end{flushleft}
\newpage

Nonlinear effects in one-dimensional surface wave propagation over a
shallow fluid have been widely studied in the past two decades in
association with solitons\cite{solitons}. What happens in two dimensions
is, by comparison, far less understood\cite{2d}. The effect of
vorticity\cite{abrash} and of submerged bodies\cite{baker} on free
surface motion has also been studied, again in one dimension,
and the vortex-surface interaction is the subject of much
current research\cite{vsi}

In a different vein, in recent years it has become apparent that
the nonlinear interaction
between sound and vorticity at low Mach numbers can be profitably
understood in terms of concepts borrowed from classical field theory of
sources and waves\cite{sourcesandwaves}.
This has allowed for an understanding of vortex dynamics in a slightly
compressible fluid\cite{vordin} and has suggested the use of ultrasound
as a probe of vorticity in both ordered and disordered flows\cite{uls}.
Recent experimental results have confirmed the soundness of this
proposal\cite{exp}. On the other hand, it is well known\cite{sound-surf}
that the nonlinear propagation of sound obeys similar equations as the
nonlinear propagation of surface waves in shallow water, and it is the
purpose of this paper to apply the concepts that have been of use to
understand the interaction of sound with vorticity to the understanding
of the interaction of two-dimensional
surface waves with ordered or disorderd vorticity in shallow water.

Consider an incompressible fluid of undisturbed uniform depth $h$
 moving with velocity $\vec v$ in a
uniform gravitational field $g$. We shall refer to it as shallow water
although of course it need not be water. The free surface will be described by
\[
z = \bar{\zeta} (x,y,t)
\]
where $\bar{\zeta} = h + \zeta(x,y,t)$, and $\zeta$ is the deviation of the
surface away from the horizontal, whose typical length scales will be
supposed to be much longer than $h$, thus allowing for the neglect of
surface tension. In shallow water vertical
variations of ${\vec v}_{\perp} $
are neglected and the governing equations are
\bea
\pder{\bar{\zeta}}{t} + \nabla_{\perp} \cdot (\bar{\zeta} {\vec v}_{\perp} )
& = & 0 \nonumber \\
\pder{{\vec v}_{\perp}}{t} + ({\vec v}_{\perp} \cdot {\nabla}_{\perp} )
{\vec v}_{\perp} & = & - g \nabla_{\perp} \bar{\zeta}
\label{one}
\eea
where the subindex $ \perp$ means horizontal, or $x-y$, components and
will be omitted from here on.
Equations (\ref{one}) are supposed to be evaluated at
$z = \bar{\zeta}$, and viscosity has been neglected. The boundary
condition at the bottom is then $v_z(z=0) = 0$.
These equations are similar
to, but not identical with, the equations for a compressible, adiabatic
bulk flow and they can be combined to yield
\beq
\frac{\partial^2 \zeta}{\partial t^2} - gh \nabla^2 \zeta  =
h \nabla
\cdot (( \vec v \cdot \nabla ) \vec v )
-\frac{\partial}{\partial t} ( \nabla \cdot ( \vec v  \zeta ) ) .
\label{two}
\eeq
This is a wave equation for the surface waves with a source term due to
the nonlinear couplings. One can think then of at least two situations
of interest when there is a bounded (two dimensional) vortical
flow with typical
velocities very small compared to $\sqrt{gh}$: one is the spontaneous
 generation of surface waves by this flow, in analogy
with the aeroacoustic generation of sound by vortical flows.
The other is the scattering of surface waves by the flow.

The reason for the surface wave generation by a bounded flow is that,
although surface deformations will not be the dominant effect at the
source, far away from it they will decay, because of their wavelike
nature, like $({\rm distance})^{-1/2}$, while the two dimensional
flow associated with the source, being incompressible, will decay like
$({\rm distance})^{-1}$. A formal solution to Eqn. (\ref{two}) can be
written as a convolution of the right hand side with the Green function
for the two dimensional wave equation. If surface elevations are
neglected at the source, an acceptable approximation, the source looks
almost like the Lighthill source term\cite{Light}. The similarity is,
however, not an identity because in the present case
${\nabla}_{\perp} \cdot {\vec v}_{\perp} \ne 0$.

In the spontaneous generation of surface waves by a flow of bounded (in
space) vorticity the second term  in the right-hand-side can safely be
neglected because surface deformations are supposed to be negligible at
the source and one obtains the following expression for the far field
surface waves spontaneously radiated by such a flow:
\[
\zeta = \frac{h}{c^2} (\vec{\omega} \wedge \vec u ) * \nabla G
\]
where $\vec u$ is a perfectly two dimensional incompressible flow
and $\vec{\omega} = \nabla \wedge \vec u $ is its (two-dimensional)
vorticity.
This is in very close analogy to the formulae for aeroacoustic sound
derived by Powell and Howe\cite{PowellHowe} which are best manipulated
in Fourier space\cite{caflund}, where
\[
G(\vec x , \nu ) = \frac{i}{8\pi} H_0^+ \left( \frac{\nu |\vec x |}{c}
\right)
\]
is a Hankel function with outgoing wave boundary condition.

The case of scattering of a plane wave can be studied with ease in the
case that wave frequencies $\nu$ are high compared to any frequencies
$\Omega$
associated with the target flow, and fluid velocities ${\vec v}_s$
associated with
the surface wave are small in comparison with those associated with the
vortical flow, $\vec u$, which in turn are supposed to be small compared with
$c \equiv \sqrt{gh}$. To this end we write
\beq
\vec v = {\vec v}_s + \vec u
\label{decomp}
\eeq
where $\vec u$ is a perfectly two dimensional incompressible vortical flow
with a perfectly flat surface and ${\vec v}_s$ ($v_s \ll u$)
is what is needed for $\vec v$ to satisfy Eqns.(\ref{one}). Moreover, the
differential equation (\ref{two}) can be turned into the integral
equation
\beq
\zeta = {\zeta}_{\rm inc} + G*s
\label{three}
\eeq
where ${\zeta}_{\rm inc} $ is the incident plane surface wave and $G*s$ is a
convolution of the Green function $G$ for the wave equation with the
source
\beq
s = h \nabla
\cdot (( \vec v \cdot \nabla ) \vec v )
-\frac{\partial}{\partial t} ( \nabla \cdot ( \vec v  \zeta ) ) .
\label{source}
\eeq

Substitution of the decomposition (\ref{decomp}) into the source
(\ref{source}) and neglecting the horizontal components of the
vorticity, a valid approximation in shallow water, as well as terms
quadratic in $v_s$, leads to a source that is the sum of three components:
\[
s= s_1 +s_2 + s_3
\]
where
\bea
s_1 & = & h \nabla \cdot ( \vec{\omega} \times {\vec v}_s ) \nonumber \\
s_2 & = & h ( \nabla^2 (\vec u \cdot {\vec v}_s ) -
\frac{\partial}{\partial t} (\vec u \cdot \nabla ) \zeta ) \nonumber \\
s_3 & = & h \nabla \cdot ((\vec u \cdot \nabla ) \vec u )    .  \nonumber
\eea

The source $s_3$ corresponds to the spontaneous generation of surface
waves discussed above
and it will not be considered further in the scattering context
since the waves so generated are
 of much lower frequency than the incoming and scattered waves.
For weak surface waves the scattered amplitude will be much weaker than
the incident one and Equation (\ref{three}) can be solved in a first
Born approximation, that is replacing $v_s$ by the incident plane wave
value
\[
{\vec v}_{\rm inc} = v_0 \hat n \cos ({\vec k}_0 \cdot \vec x - \nu_0 t)
, \]
for which velocity at the surface and surface elevation are related by
\[
\pder{\vec v}{t} = -g \nabla \zeta .
\]
Using the fact that, in the far field
\[
\nabla G \approx -\frac{\hat x}{c} \pder{G}{t}
\]
and integrating by parts it is tedious but straightforward to show that
if the scattered wave is written as
\[
\zeta_{\rm scatt} = \zeta_1 + \zeta_2
\]
where $\zeta_a = G*s_a$, $a=1,2$, then
\[
\zeta_1 = ({\hat k}_0 \cdot \hat x -1 ) \zeta_2 + \frac{1}{g c^2}
G*\left( \pder{\vec u}{t} \cdot \pder{{\vec v}_{\rm inc}}{t} -{\hat k}_0 \cdot
\hat x \frac{\partial}{\partial t} \left( \pder{\vec u}{t} \cdot
{\vec v}_{\rm inc} \right) \right) .
\]
The second term on the right is of order $\sim \Omega \nu u v_0 /c^2$ and can
be neglected with respect to $\zeta_2$, which is of order $\sim k_0^2 u
v_0$, in the case under consideration of incident frequencies much
higher than typical frequencies associated with the vortical target:
$\Omega \ll \nu_0$.

We are thus finally led to the relation
\beq
\zeta_{\rm scatt} = \frac{\cos \theta}{g (\cos \theta -1)} G* \left(
\nabla \cdot (\vec{\omega} \times {\vec v}_{\rm inc} ) \right)
\label{four}
\eeq
where $\theta$ is the scattering angle and $\vec{\omega}$ the vorticity.
Using the far-field expression for $G$,
\[
G(\vec x , \nu) \sim \frac{1}{4\pi} \left( \frac{c}{2\pi \nu |\vec x | }
\right)^{1/2} \exp i ( \frac{\nu |\vec x |}{c} + \frac{\pi}{4} )
\]
 this expression becomes
\beq
\zeta_{\rm scatt} (\vec x , \nu) = h \left( \frac{\sin \theta \cos
\theta}{\cos \theta -1} \right) \left( \frac{\pi^3 \nu}{2 c^3 |x|}
\right)^{1/2} \left( \frac{v_0}{c} \right) e^{i(\nu |x| /c + 3\pi /4) }
\hat{\omega} (\vec q , \nu - \nu_0)
\label{five}
\eeq
which is the promised relation between scattered surface wave amplitude
at position $\vec x$ and frequency $\nu$ in terms of $\hat{\omega}$,
the Fourier transform of vorticity. The incident wave has (velocity)
amplitude $v_0$ and frequency $\nu_0$; $\vec q$ is the momentum
transfer. Note that the apparent divergence of this expression for small
scattering angles is an artifact of taking an incident plane wave which
is, technically, infinite. The angular dependence will break down for
wavelengths $\lambda$ at angles such that
\[
\sin \theta  \sim \lambda /L
\]
where $L$ is a length of the order of the spatial extent of the incident
wave.
Formula (\ref{five}) suggests a possible non intrusive way to
probe vortical flows in shallow water using surface wave scattering.

Berry and collaborators\cite{berry} have remarked that a plane wave of
wave vector $\vec k$
propagating along the $x$ direction
incident on a stationary point vortex will give rise to surface deformations
that can be locally (as opposed to globally) described by a multivalued
phase:
 \[
 \zeta \sim e^{i(kr\cos \theta +\nu t +\alpha \theta)}
 \]
 corresponding to wavefronts exhibiting a dislocation at the position of
 the vortex. Here
  $(r,\theta)$ are polar coordinates centered at the vortex and
 $\alpha = \nu \Gamma /c^2 $, where $\Gamma$ is the vortex circulation.
 For the dislocation to be noticeable $\alpha$ must be of order one.
 It is of interest to see how this is related to the results reported
 above, which are not immediately applicable since the scattering of
 surface waves by vortical flows was worked out for high frequency
 waves, for which this last condition is violated. Indeed, at a distance
 $\sim \lambda$ from a point vortex the frequency of the vortical flow
 is $\Omega \sim \Gamma / \lambda^2 $ which is $\sim \nu$ when $\alpha
 \sim 1$.

In the presence of a steady divergenceless
background flow $\vec U$ giving rise
to surface deformations $\zeta_0$, small velocity $\vec v$
and surface deviations $\zeta$
will obey the equations, easily derived from (\ref{one}),
\bea
\left( \pder{}{t} + \vec U \cdot \nabla \right ) \zeta + h \nabla^2
\tilde \phi & = & 0  \nonumber \\
\left( \pder{}{t} + \vec U \cdot \nabla \right ) \tilde \phi + g \zeta &
= & 0
\label{back}
\eea
where $\tilde \phi$ is the velocity potential for the velocity, $\vec v =
\nabla \tilde \phi$, which exists locally outside the vortex core.
Quadratically small terms have been neglected and position
is supposed to be sufficiently far away from the vortex core that
gradients are dominated by wavevectors. Under the additional assumption
that $U^2 \ll c^2$, equations (\ref{back}) yield
\beq
\frac{\partial^2 \tilde \phi}{\partial t^2} + 2 \vec U \cdot \nabla
\pder{\tilde \phi}{t} -c^2 \nabla^2 \tilde \phi = 0 .
\eeq
Note now that the velocity field outside a vortex is the gradient of a
multivalued scalar potential which is proportional to the polar angle
centered at the vortex:
\[
\vec U = \nabla \left( \frac{\Gamma \theta}{2 \pi} \right)  .
\]
Looking now for time harmonic solutions in the form
\[
\tilde \phi = \phi e^{i\nu t} e^{i \Phi}
\]
we have that with
\[
\Phi = \frac{\nu \Gamma \theta}{2 \pi c^2}
\]
$\phi$ obeys the equation
\[
(\nabla^2 + \frac{\nu^2}{c^2} ) \phi = 0 ,
\]
giving rise to the dislocated wavefronts of Berry et. al.\cite{berry}.

In addition to the dislocated incident wave there is a
scattered wave. In the case of the Aharonov-Bohm effect\cite{berry} the
latter can be calculated using vanishing boundary conditions at the ``vortex''
core. The scattering mechanism in the fluid case is different, its
origin being not in impenetrable boundary conditions but in the
nonlinear interaction terms that were neglected above. The computation
of this effect offers an interesting challenge which is, however,
outside the scope of the present paper.

To conclude, we have studied the nonlinear interaction between surface
waves (for a two dimensional surface) and vorticity in shallow water.
Analytical formulae for the generation and scattering of such waves by
any (two-dimensional) vortical flow have been derived under the following
assumptions: a) Time scales are such that viscosity can be neglected; b)
Length scales are such that surface tension can be neglected; c)
For scattering, the frequency of the incident wave is high by comparison
with the (inverse of) the time scale of the vortical flow, and the
particle velocity associated with the wave is supposed to be small
compared with the velocity of the vortical flow, which in turn is
supposed to be small compared to $\sqrt{gh}$, the phase velocity of the
waves. Dislocated wavefronts analogous to the Aharonov-Bohm effect have
been obtained as a special case. The question naturally arises as to
what happens when the fluid is not shallow. The point of view presented
in this paper, namely that of studying the nonlinear interaction between
surface motions and vorticity by way of succesive approximations using
ideas borrowed from classical field theory, may be
of use in studying this problem.

This work was supported by DTI Grant E2854-9244, FONDECYT Grant
1265-91 and CEC Contract CI1*-CT91-0947.

\end{document}